# An New Type Of Artificial Brain Using Controlled Neurons

John Robert Burger, Emeritus Professor, ECE Department, California State University Northridge,
jrburger1@gmail.com

***Abstract --*** Plans for a new type of artificial brain are possible because of realistic neurons in logically structured arrays of controlled toggles, one toggle per neuron. Controlled toggles can be made to compute, in parallel, parameters of critical importance for each of several complex images recalled from associative long term memory. Controlled toggles are shown below to amount to a new type of neural network that supports autonomous behavior and action.

## I. INTRODUCTION

A brain-inspired system is proposed below in order to demonstrate the utility of electrically realistic neurons as controlled toggles. Plans for artificial brains are certainly not new [1, 2]. As in other such systems, the system below will monitor external information, analogous to that flowing from eyes, ears and other sensors. If the information is significant, for instance, bright spots, or loud passages, or if identical information is presented several times, it will automatically be committed to associative long term memory; it will automatically be memorized.

External information is concurrently impressed on a short term memory, which is a logical row of short term memory neurons. From here cues are derived and held long enough to accomplish a search of associative long term memory. Unless the cues are very exacting indeed, several returned images are expected. Each return then undergoes calculations for priority; an image with maximum priority is imposed on (and overwrites the current contents of) short term memory. Calculations for priority are fast because they proceed in parallel, which is one of the purposes of an array of controlled toggles.

Short term memory holds images alternating between external information, and recalls from long term memory. Each image in short term memory causes a search of associative long term memory, with a resulting recall, as well as physical actions. The system to be described is autonomous and follows from neuro-cognitive psychology studies.

Little attention is paid below to classical artificial neural networks, even though they have potential to serve admirably in the detection of given attributes hidden in sensor data. What is needed, but elusive, is massive parallel processing, to enable fast decisions concerning the priority of recalled images. To address this challenge, the author found it necessary to ignore classical artificial neural networks, as well as those elegant mathematical theories that in fact are irrelevant to the fast computation of recall priorities.

A first step in departing from older, inappropriate technologies is to go back to the logical functions of an actual biological neuron. Neurons use pulses instead of voltage levels. With pulses, any Boolean function is readily available [3-10]. The neural membrane that supports pulsing is fairly efficient, since pulses tend to be slow and physically reversible; they tend to leave the system the way they found it [3, 11]. A simple neuron can serve as a toggle, and with a minor adjustment, a controlled toggle [12].

A controlled toggle changes its state only if control signals from one or more other toggles all are true. The important thing about neural toggles is that they may be organized into several logical rows, each holding a given image, each processing in parallel a given set of instructions. Parallelism in logically reversible systems based on controlled toggles is very natural [3, 4, 13]. A wide variety of mathematical calculations are readily possible, although this paper is concerned only with a calculation of priority for each recalled image.

The humble neuron accomplishes more logic in a smaller space, using less energy than most integrated solid state devices. Artificial electrically realistic neurons have yet to be perfected for engineering purposes. Meanwhile, transistors, albeit inadequate, may serve to implement the controlled toggles and other logic necessary to the system described below.

## II. FRONT END

An autonomous memory machine (or artificial brain) serves to illustrate the utility of electrically realistic neurons in the form of controlled toggles. Figure 1 shows a plan created from previous efforts [3, 4]. Sensory information, which is a maze of signals, undergo, in Block 1, decoding into attributes, or basic elements of edges, shades, colors, tone, and so on. Block 1 is not discussed here, except to note that it may very well contain classical neural network structures. Attributes are assumed to be digital; a pulsating neural output means TRUE, no signal means FALSE.

Attributes, K in number, are used in three ways: 1) they provide stimulation to provide special displays of the environment. 2) they provide "key" attributes that serve to detect unusual "events" or repetition of an "event" as in Blocks 2, 3. If this event is not already memorized, meaning there is a NO HIT after a memory search, described next, then there is issued an ENABLE MEMORIZATION signal from gate 4. This means that the event, or image, is automatically memorized in available unused memory; as is described elsewhere [3, 4, 13].

Attributes are also impressed on short term memory, Block 5, where certain major attributes provide cues for a memory search. Normally, cues are under determined. Thus associative memory provides several images, and a HIT signal is emitted; returned images are subsequently evaluated for priority (explained below).

However, in the rare cases when there is a NO HIT signal, a Cue Editor, Block 6, is helpful to assure that images are found. The Cue Editor randomly removes attributes, making a HIT more likely [3, 4]. The Cue Editor works in the background, analogous to when a person has a mental block, but remembers later.

A simple subsystem alternates what is impressed on short term memory, either sensory attributes or recalled images, both of which result in additional memory searches. The subsystem is a single neuron, Block 7 whose output axon synapses back to its own dendrite, with suitable delay, probably tens of milliseconds in a human brain. Unit 8, a bundle of logical neurons, serves as a 2x1 multiplexier.

Note that recalled images, or impressions in short term memory, are perceived only dimly by humans; they are not as bright or loud as the original simulation. But they are still perceived, and may result in important physical actions, denoted by Block 9. Short term memory, anthropomorphically, is a center of "consciousness."

## III. IMAGE PRIORITIES

The system and neural circuitry of associative memory, as in a brain, have been analyzed elsewhere [4, 13]. After a successful search, and the occurrence of a HIT signal, images flow asynchronously, and are effortlessly directed into registers of toggle neurons as suggested in the top of Figure 2. The load pulse, by the way, is derived from the HIT signal, which is assumed converted into a single membrane-like pulse. "Importance" attributes, κ in number, a small subset of all attributes, are immediately converted into binary weights each $N_1$ bits, termed "subpriorities." They can be converted by direct connections. Immediate physical danger would receive the highest value, emotional events slightly less, loud noises, or flashes of light even less, and so on to a value of zero [1, 2].

The bottom of Figure 2 illustrates toggle registers, amounting to a number L of neural controlled toggles. Within these registers calculations of priority are to be computed. These bottom registers will each require extra scratchpad toggles.

The calculation involves adding all subpriority codes for each register in parallel. Each register has its own adder implemented in scratchpad toggles, as delineated elsewhere [3, 4, 13]. Figure 3 illustrates that within each register the codes are added two at a time. This is done for all registers in parallel, using κ − 1 adder blocks. Adder operations are controlled by a sequence of "source" and "target" signals flowing asynchronously from a long term memory (LTM).

Note in Figure 3 that the A-inputs serve to hold an accumulation sum; the B-inputs sequentially add in the other subpriority codes ($X_i$, $3 \leq i \leq \kappa$). The priority of each image resides in the summation outputs.

A controlled toggle structure is portrayed in Figure 4. If the connecting bus is at rest then the toggle neuron $Q_i$ will have its state flipped if there is a Target signal from LTM. If there is a Source signal from LTM, and if the toggle $Q_i$ is TRUE, with pulsing output, the bus remains at rest. But if $Q_i$ is FALSE, then pulses are transmitted to the bus, meaning no bus toggle can be flipped. This system permits the testing of several toggles, and only if they are all TRUE will the bus remain at rest, with the possibility of toggling those other toggles whose Target signal is active (pulsating). This system permits a variety of complex computations.

Using hardware, a single wire would serve as the bus as in Figure 5. But biologically, an interneuron is required with many dendritic synaptic inputs and many axon-like buttons. Since neurons are so very plentiful in a brain, almost a trillion ($10^{12}$), interneurons are readily available.

Figure 6 illustrates comparing magnitudes of computed priorities and using them to direct a given image, the highest priority, into short term memory. Magnitude comparison may be accomplished using reversible logic in which net priorities are subtracted to determine the largest, or alternately, using a combinational magnitude comparator. Nearly everything in this plan is asynchronous, with no central clock. Timing has to be implemented correctly, although it develops naturally in a biological brain.

## IV. CONCLUSIONS

This article suggests a new direction for artificial neural networks, using structures that differ radically from the usual tree of many weighted inputs, connected by nonlinear amplifiers with a few logical outputs. There are no weights, especially no synaptic weights in the above concept; a neural membrane is either triggered, or it is not. There are no negative numbers either, although there are excitatory and inhibitory synapses.

Controlled toggles using biological neurons are energy efficient, occupy a small space, and serve well for parallel computations. To grasp this from the bottom up, one must study the given references. The future may well reside in artificial pulsating neurons for controlled toggles, assuming such neurons can someday be manufactured. This paper allows a reader to comprehend a particular artificial brain, one that achieves alertness, has a low level of intelligence, and rudimentary self-protection, all of which depend on the availability of controlled toggles.

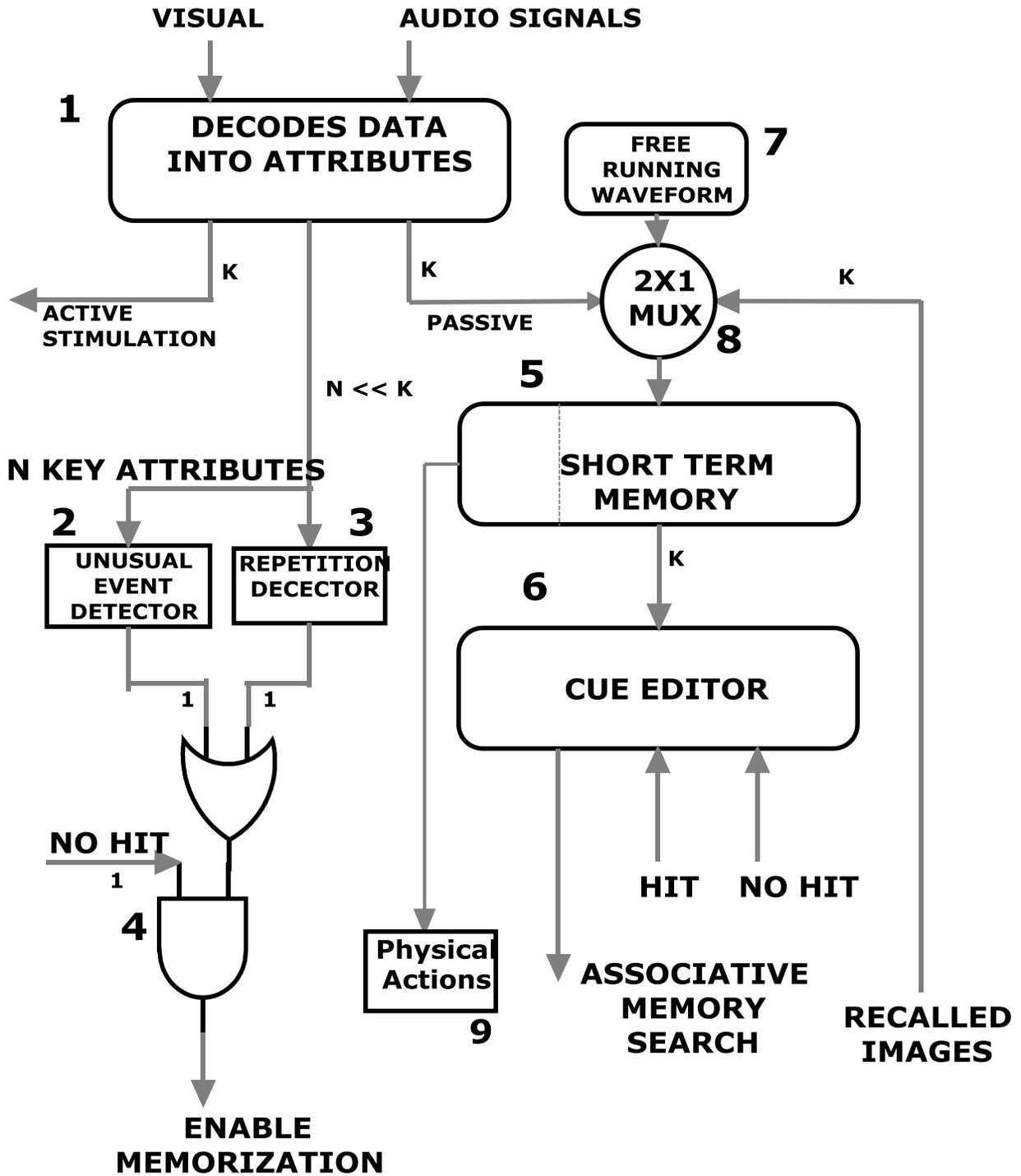

**Figure 1. Artificial Brain Front End.**

New Developments in Neural Networks, *J. R. Burger*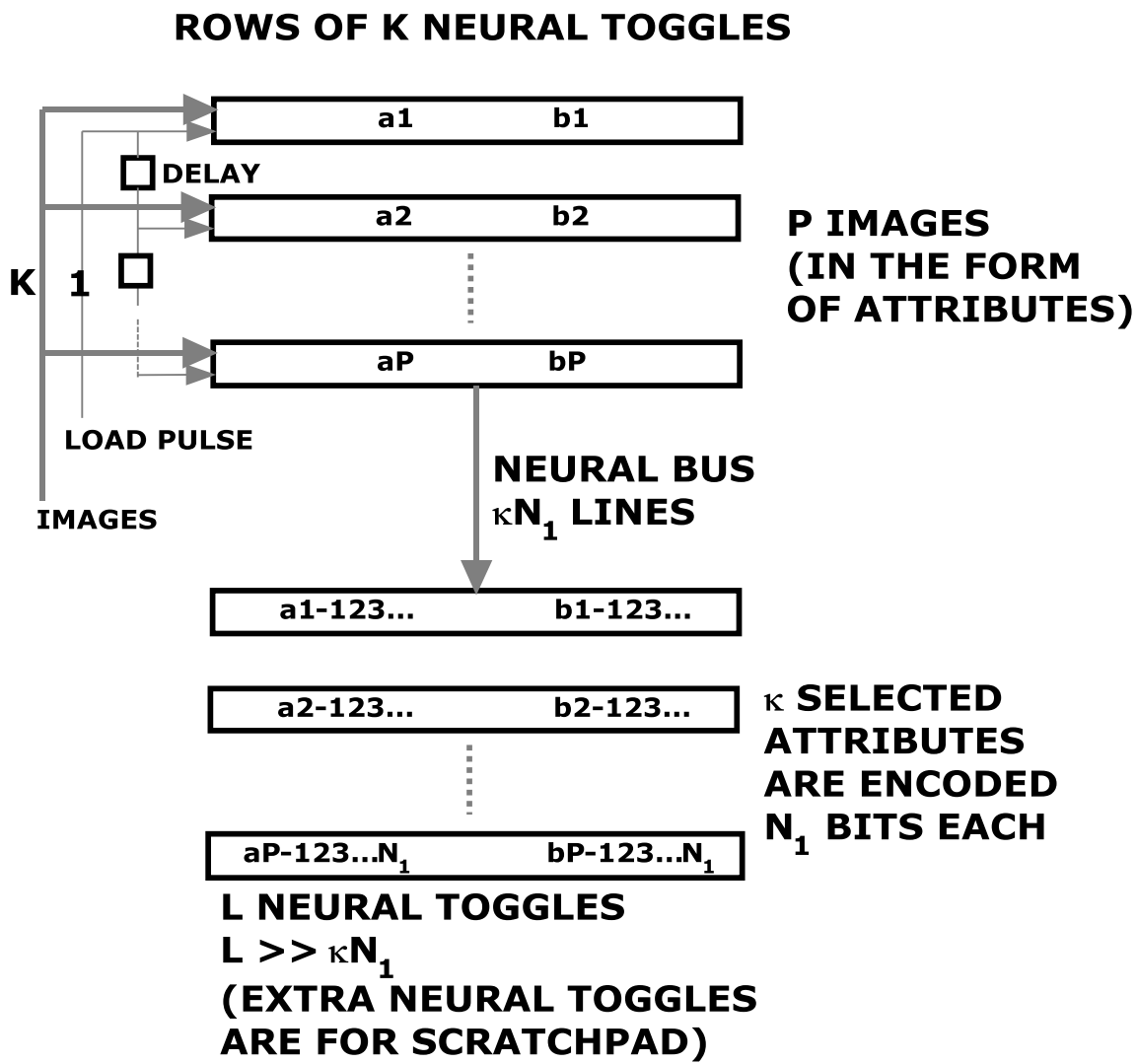

Figure 2.   Registers of Controlled Toggles.

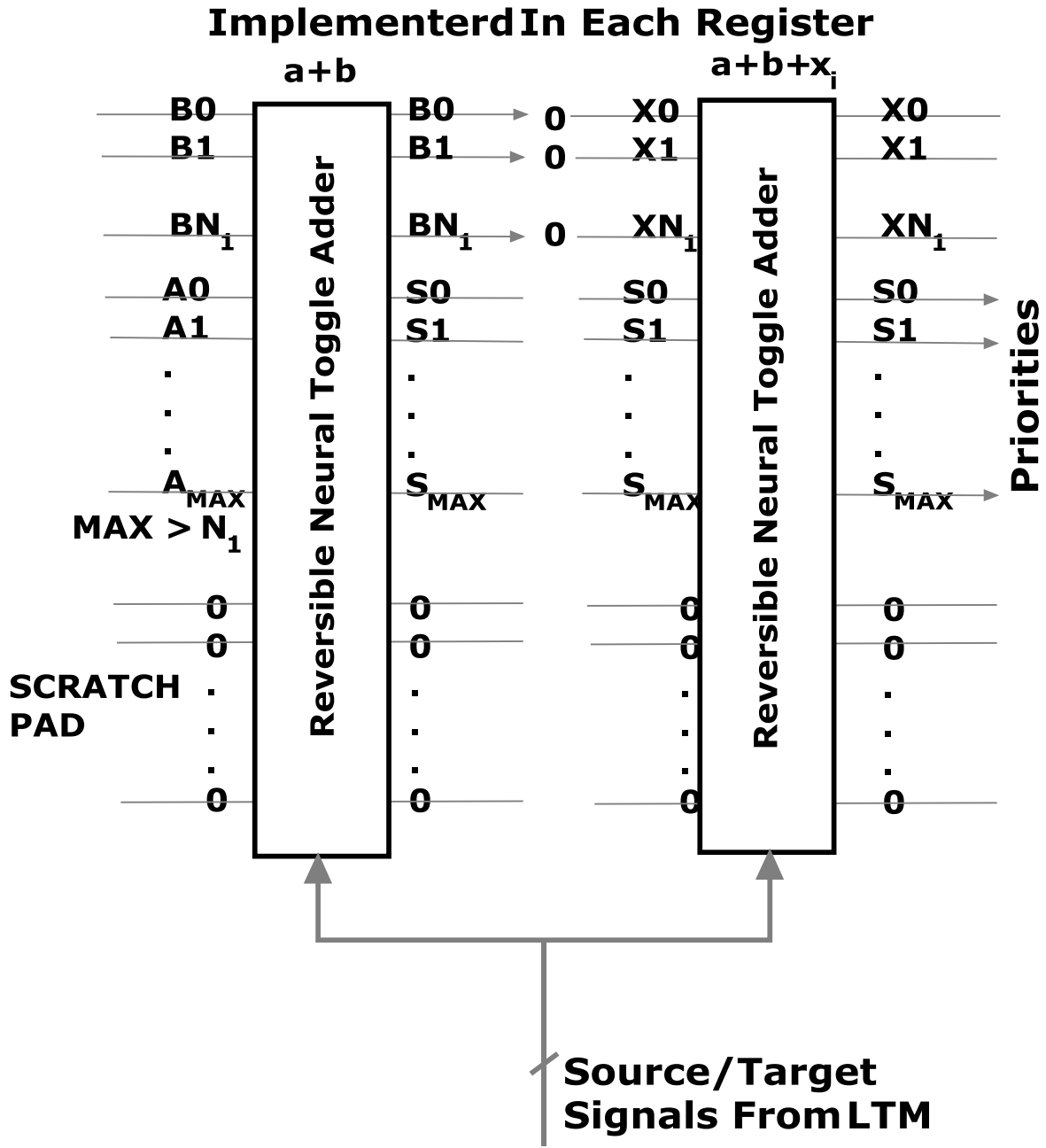

**Figure 3. Addition Within A Toggle Register.**

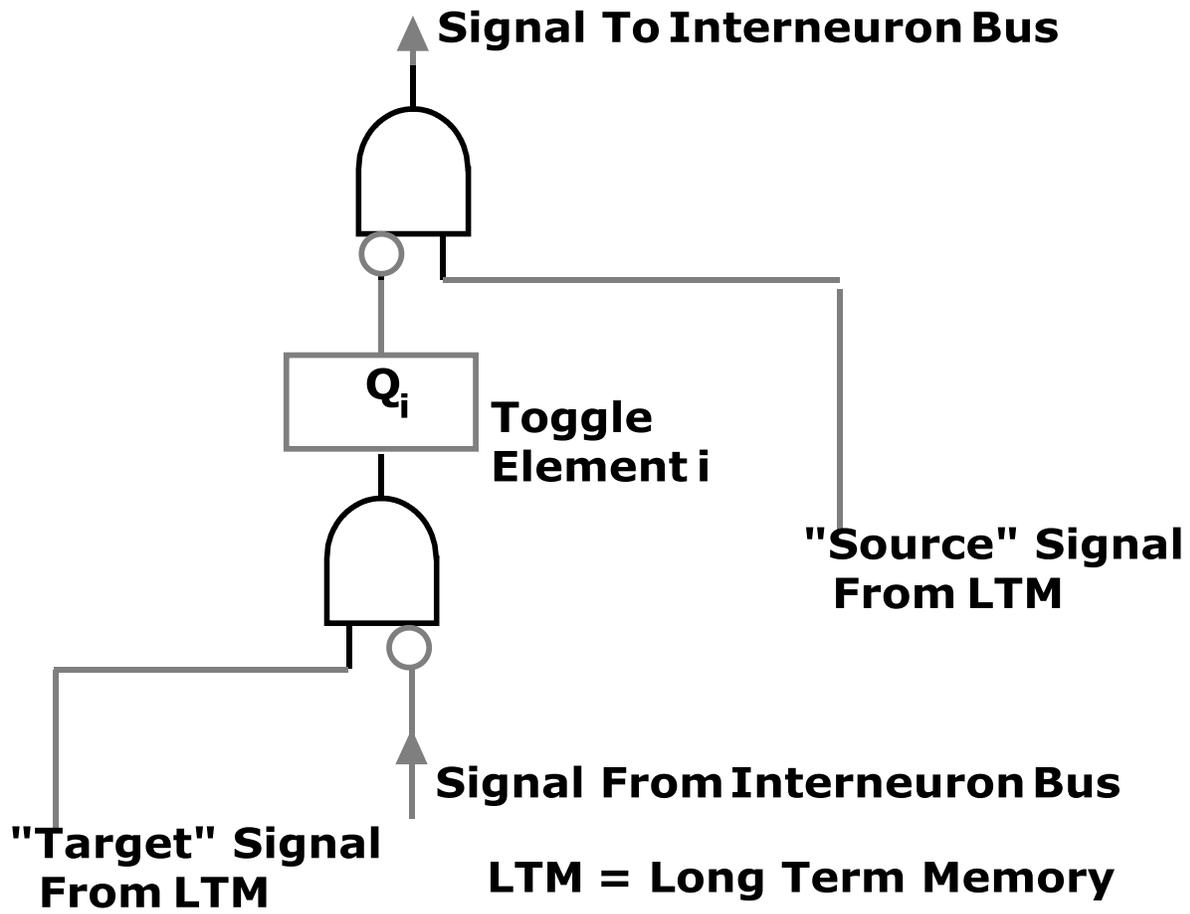

**Figure 4.** Controlled Toggle Element.

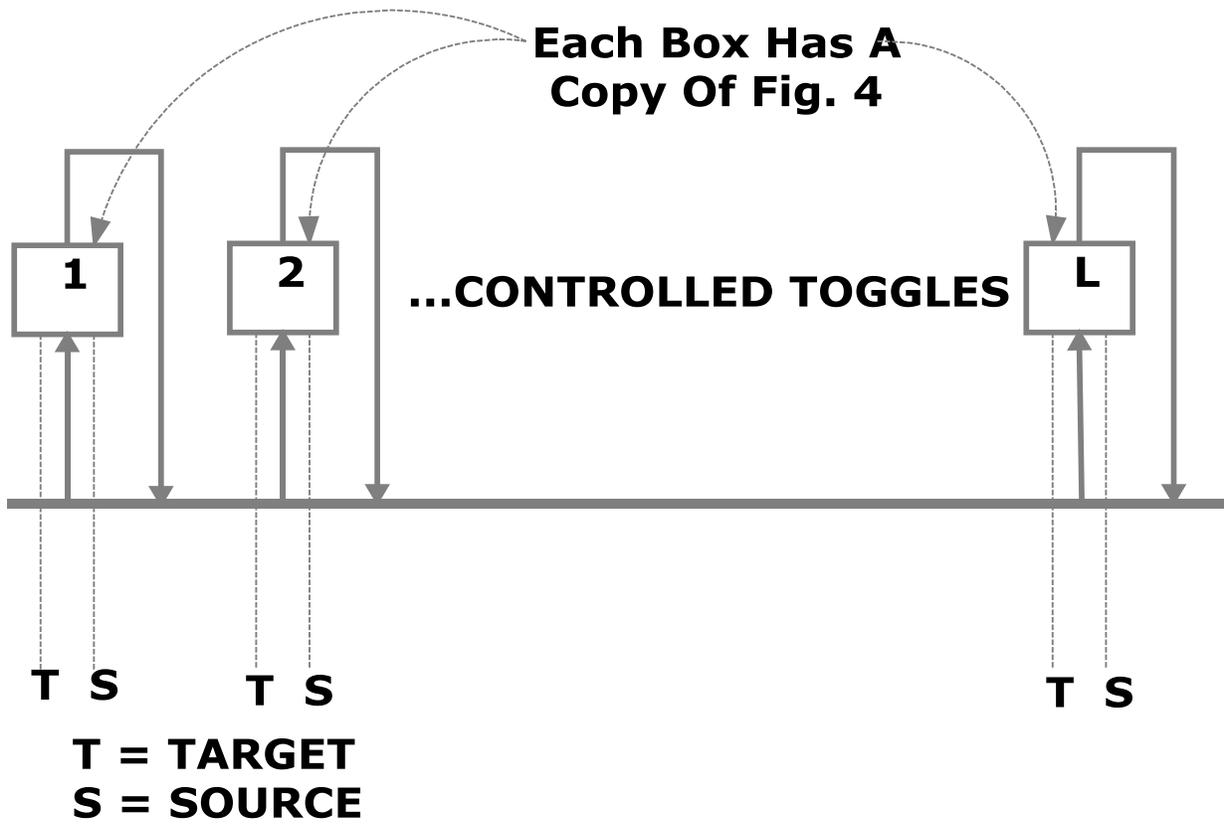

**Figure 5.   A Neural Bus System.**

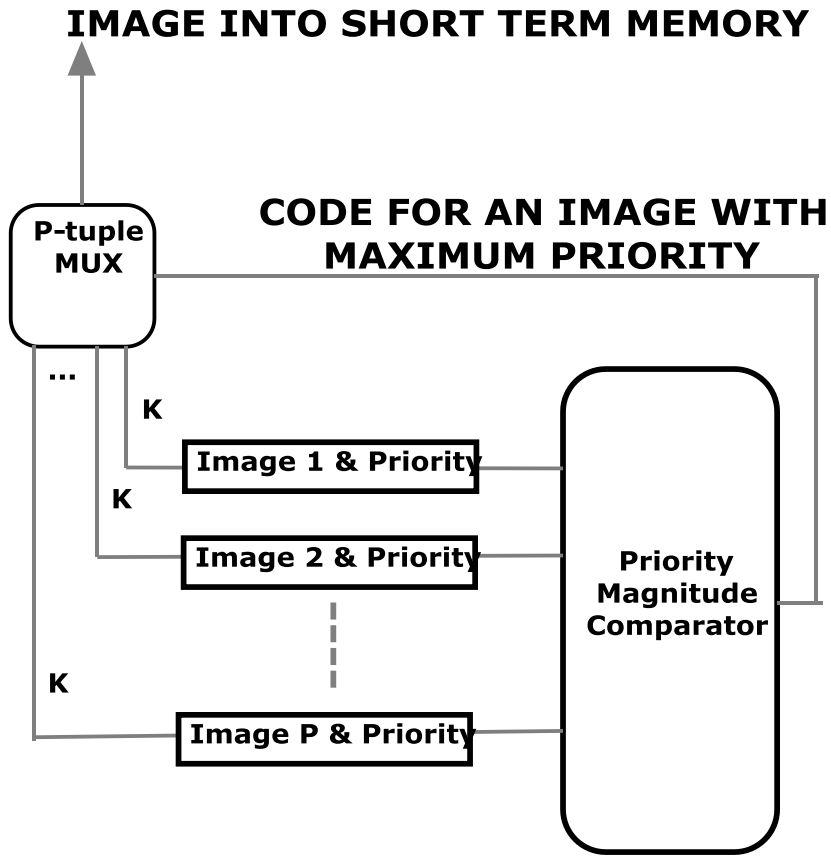

Figure 6.   Selecting The Highest Priority Image.